\documentclass[aps,prb,twocolumn]{revtex4-1}
\pdfoutput=1
\usepackage{amsmath}
\usepackage{ amssymb }
\usepackage{graphicx}
\usepackage{color}

\begin{document}

\newcommand{\beq}{\begin{equation}}
\newcommand{\eeq}{\end{equation}}
\newcommand{\beqa}{\begin{eqnarray}}
\newcommand{\eeqa}{\end{eqnarray}}
\newcommand{\note}[1]{{\color{red} [#1]}}
\newcommand{\bra}[1]{\ensuremath{\langle#1|}}
\newcommand{\ket}[1]{\ensuremath{|#1\rangle}}
\newcommand{\bracket}[2]{\ensuremath{\langle#1|#2\rangle}}
\renewcommand{\vec}[1]{\textbf{#1}}
\newcommand{\dagga}{{\phantom{\dagger}}}


\title{Thermal Ising transitions in the vicinity of two-dimensional quantum critical points}

\author{S. Hesselmann}
\affiliation{Institut f\"ur Theoretische Festk\"orperphysik, JARA-FIT and JARA-HPC, RWTH Aachen University, 52056 Aachen, Germany}

\author{S. Wessel}
\affiliation{Institut f\"ur Theoretische Festk\"orperphysik, JARA-FIT and JARA-HPC, RWTH Aachen University, 52056 Aachen, Germany}

\date{\today}

\begin{abstract}

The scaling of the transition temperature into an ordered phase close to  a quantum critical point as well as the order parameter  fluctuations inside the quantum critical region provide valuable information about  universal properties of the underlying quantum critical point. Here, we employ quantum Monte Carlo simulations to examine these relations 
in detail for two-dimensional quantum systems that exhibit a finite-temperature Ising-transition line in  the vicinity of a quantum critical point that belongs to the universality class of either (i) the three-dimensional Ising model for the case of the quantum Ising model in a transverse magnetic field on the square lattice or (ii) the  {\it chiral} Ising transition for the case of a half-filled system of spinless fermions on the honeycomb lattice with nearest-neighbor repulsion. While  the first case allows large-scale simulations to assess the scaling predictions to a high precision in terms of the known values for the critical exponents at the quantum critical point, 
for the later case we extract values of the critical exponents $\nu$ and $\eta$, related to  the order parameter fluctuations, which we discuss in relation to other recent estimates from ground state quantum Monte Carlo calculations as well as analytical approaches.

\end{abstract}

\maketitle

\section{Introduction}\label{sec:intro}

Quantum phase transitions in systems of interacting fermions with a relativistic  free dispersion have been  investigated intensively in recent years. Beyond their fundamental relevance in relativistic quantum field theory, such fermion systems also emerge in the low-energy sector of  various condensed-matter systems such as for electrons on graphene's honeycomb lattice~\cite{CastroNeto09}, ultracold fermions in optical lattices~\cite{Uehlinger13}, $d$-wave superconductors~\cite{Vojta00, Khveshchenko01}, and surface states of topological insulators~\cite{Hasan10}. 
In case the density of states  vanishes at the free system's Fermi energy, 
effectively relativistic fermion systems are robust to weak interactions, and a finite critical interaction strength is required to drive an instability towards a Mott-insulating phase, wherein the fermions acquire a finite mass from chiral symmetry breaking. 
For lattice fermions, 
this is the case at  specific commensurate fillings, such that a low-coupling semi-metallic phase is separated from an insulating phase by a finite-coupling quantum critical point. 

Along with the chiral symmetry breaking, 
the Mott-insulating ground state may also exhibit long-range order such as in an antiferromagnetic or charge density wave (CDW) state. 
Due to the coupling of the order parameter fluctuations to  low-energy fermion excitations, the universal behavior at the corresponding quantum critical point differs  from the one expected from a naive analysis in terms of the number $N$ of components of the order parameter field and the system's dimensionality: a prominent example is provided by the critical Gross-Neveu-Yukawa theory~\cite{Gross74, Herbut06, Herbut09a} for the quantum critical point of a system of $N_f$ flavors of relativistic four-component Dirac fermions with a coupling to an $N$-component  order-parameter field. The scaling properties, such as critical exponents, have been investigated in the past by various analytical and numerical methods, and have been established to be distinct from, e.g.,  those of critical points captured by the conventional (classical) $O(N)$ symmetric  $\phi^4$-theory~\cite{Rosenstein93,Rosa01,Hoefling02,Mesterhazy12,Vacca15}.
These distinct universality classes for chiral symmetry breaking are also  referred to as {\it chiral}, e.g., for $N=1$ as the chiral Ising (${Z}_2$) universality class, and correspondingly for larger values of $N$ (e.g., chiral Heisenberg for $N=3$). 

From a condensed-matter perspective on two-dimensional quantum lattice systems, 
the $N=1$ case of the $Z_2$-Gross-Neveu theory exhibits a further interesting aspect, since in addition
to the chiral Ising quantum critical point, the discrete symmetry of the system then allows for an extended symmetry-broken phase also at finite temperatures,  terminated by  a line of finite-temperature phase transitions that restores the chiral symmetry. 
Anticipating the decoupling of the fermions from the  critical order parameter fluctuations at finite temperatures, 
the finite-temperature transitions belong to the universality class of the classical two-dimensional Ising model, in accord with  the general principles of dimensional reduction and universality~\cite{Stephanov95}.
Enhanced  fluctuations   however drive the ordering temperature to zero upon approaching the quantum critical coupling strength.
Based on general scaling considerations within the scaling regime  of the quantum critical point~\cite{Sachdev11},  the  Ising transition temperature $T_c$ is furthermore expected to scale with the detuning of the dimensionless interaction strength $g$ from its quantum critical value $g_c$ as 
\beq\label{eq:Tcvsg}
T_c\propto |g-g_c|^{z\nu},
\eeq
describing the termination of the transition temperature at the quantum critical point, and 
where $z$ denotes the dynamical critical exponent and $\nu$ the correlation length exponent for the order parameter fluctuations of the underlying quantum critical point. This relation thus connects directly the finite-temperature  phase boundary line near the $T=0$ quantum critical point to its  quantum critical exponents~\cite{footnote1}.

While relativistic invariance locks the value of $z=1$, other critical exponents, including $\nu$, are less precisely established for the chiral Ising universality class. In the past, estimates for these critical exponents were obtained from approximate renormalization group calculations~\cite{Rosenstein93,Rosa01,Hoefling02,Mesterhazy12,Janssen14,Vacca15}, and more recently have been extracted also from quantum Monte Carlo (QMC) simulations of appropriate fermionic quantum lattice models~\cite{Wang14a,Li15b,Wang15a}. 
We will review and discuss these various estimates for the critical exponents in Sec.~\ref{sec:discussion}.
A particular useful lattice-based regularization of the $N_f=1$, $Z_2$-Gross-Neveu theory in 2+1 dimensions is supposed to be provided by a half-filled system of spinless fermions on the honeycomb lattice, with a nearest-neighbor hopping amplitude $t$ and interaction strength $V>V_c$.
Indeed, only recently  has it been realized~\cite{Huffman14,Wang14a}, that such a system can be studied by unbiased and sign problem-free continuous-time QMC simulations that can probe directly the correlations across the quantum critical point that resides at a critical coupling strength $V_c$, and which separates a low-$V$ semi-metal from the CDW state at large values of $V$.
In addition to the continuous-time interaction expansion (CT-INT) algorithm of Ref.~\onlinecite{Huffman14,Wang14a}, 
also a QMC algorithm based on a Majorana formulation~\cite{Li15a,Li15b} (MQMC) and a projective
continuous-time approach~\cite{Iazzi15,Wang15a} (LCT-INT) have been applied to this model, yielding consistent findings. More recently,  close connections among these algorithmic approaches have furthermore been identified~\cite{Wang15b,Wei16,Li16}.
From these recent QMC simulations, which concentrated on ground state properties, the value of the quantum critical interaction strength $V_c\approx 1.355 t$ in units of the hopping strength has been estimated, and approximate values of the critical exponent $\nu$ and the anomalous exponent $\eta$ for the order parameter fluctuations have been obtained, which we review in more detail below.
Furthermore, the location of the quantum critical point was also confirmed from analyzing the 
scaling of entanglement measures in this model~\cite{Broecker15}. 
 
In the following, we employ the  CT-INT approach to examine the spinless fermion $t-V$ model on the honeycomb lattice in more detail at {\it finite} temperatures. In particular, we determine the thermal Ising transition line in this model and the critical exponent $\nu$ from the relation in Eq.~(\ref{eq:Tcvsg}), as well as the anomalous exponent $\eta$. The later is estimated from performing finite-temperature simulations within the quantum critical regime atop the quantum critical point. 
In addition to the fermionic $t-V$ model, we furthermore consider the quantum spin (transverse-field) Ising model on the square lattice, which also features a finite-temperature Ising transition  terminating at a quantum critical point~\cite{Sachdev11}. However, in this case,  the quantum critical point belongs to the universality class of  the three-dimensional classical Ising model, with rather accurately established values for the critical exponents. This fact allows us to examine the asymptotic scaling form in Eq.~(\ref{eq:Tcvsg}) in more detail, in particular since  we can employ in this case large-scale QMC simulations based on the stochastic series expansion (SSE)~\cite{Sandvik03}. For the $t-V$ model, the system sizes accessible by the CT-INT method are more restricted, due to the cubic scaling of the algorithmic complexity with the overall system size (while in the SSE, the numerical effort scales linearly with the system size). 

The further organization of this paper is obvious from the section headings: in Sec.~\ref{sec:model}, we present in more detail the models and the numerical methods for  our analysis. We then present in  Sec.~\ref{sec:ising} our results for the finite temperature scaling behavior near the quantum critical point for the quantum Ising model, and in Sec.~\ref{sec:tV}, we
present our results for the $t-V$ model on the honeycomb lattice and compare to the values of the critical exponents from  previous ground state QMC simulations.  Section ~\ref{sec:discussion} contains our final conclusions.

\section{Models and methods}~\label{sec:model}
The model of primary interest to our investigations is the $t-V$ model of spinless fermions on the honeycomb lattice, described by the  Hamiltonian
\beqa
\label{eq:HtV}
H&=&H_0+H_I,\\
H_0&=&-t \sum_{\langle i,j \rangle} \left( c^\dagger_i c^\dagga_j + c^\dagger_j c^\dagga_i \right), \nonumber\\
H_I&=& V \sum_{\langle i,j \rangle} \left( n_i-\frac{1}{2}\right) \left( n_j-\frac{1}{2}\right), \nonumber 
\eeqa
where $c^\dagger_i$ ($c^\dagga_i$) creates (annihilates) a spinless fermion on the $i$th lattice sites, and  
 both summations extend over the set of nearest neighbor bonds of the honeycomb lattice. 
 The interaction term is written in explicit particle-hole symmetric form at half-filling, which we consider in the following. 
While for values of $V<V_c$, the half-filled system resides within a semi-metallic phase, a staggered CDW Mott-insulator state is stabilized at low temperatures for $V>V_c$. It is signalled by a finite value in the thermodynamic limit of the squared CDW order parameter estimator
\beq
\label{eq:m2tV}
M_2=\frac{1}{N_s^2}\sum_{i,j} \epsilon_i \epsilon_j \left\langle \left( n_i-\frac{1}{2}\right) \left( n_j-\frac{1}{2}\right)\right\rangle,
\eeq
where $N_s$ denotes the number of lattice sites, and $\epsilon_i$ is a binary variable that takes on the values $\pm 1$, depending on the sublattice to which the $i$th lattice site belongs, and thus accounts for the staggered oder in the CDW phase. 
We also denote by $g=V/t$ the dimensionless tuning parameter,  and $g_c$ refers to  its critical value. 

This model has been examined  in several recent works, in particular since it was realized, that it can be studied by unbiased, sign problem-free QMC methods, based either on the fermion bag approach~\cite{Chandrasekharan10, Huffman14, Wang14a} or an appropriate decoupling of the interactions~\cite{Iazzi15,Wang15a,Wang15b}, e.g., after expressing it in terms of Majorana fermions~\cite{Li15a,Li15b}. In the following, we use the continuous-time interaction expansion (CT-INT) approach presented in Refs.~\onlinecite{Huffman14,Wang14a} to study the system with CT-INT at finite temperatures, as detailed in Sec.~\ref{sec:tV}.
In particular,  we explore the finite temperature properties of this model in the vicinity of the quantum critical point, which is assumed to belong to the chiral Ising universality class, as referred to in Sec.~\ref{sec:intro}.

In order to contrast this scenario in a fermionic model to the corresponding behavior near a conventional Ising quantum critical point, we also analyze 
in the following the quantum Ising model in a transverse magnetic field with the Hamiltonian
\beq
H_Q=-J \sum_{i,j} \sigma^z_i \sigma^z_j -\Gamma \sum_i \sigma^x_i,
\eeq
of local spin-1/2 degrees of freedom, described in terms of local Pauli matrices.
Here,  we consider  for convenience the well-studied case of an underlying  square lattice geometry, with a  ferromagnetic nearest neighbor  interaction $J>0$ and the transverse field $\Gamma$. From previous studies, it is known,
that the low-$\Gamma$ ferromagnetic order is destroyed for transverse fields beyond the critical field strength of $\Gamma_c=3.04438(2) J$, where  a paramagnetic alignment of the spins in the field direction sets in, which destroys the long-ranged correlations of the low-$\Gamma$ ferromagnetic phase~\cite{Blote02}. The ferromagnetic order can be accessed in terms of a finite value in the thermodynamic limit of the squared order parameter estimator
\beq
\label{eq:m2Ising}
M^Q_2= \left\langle \left( \frac{1}{N_s} \sum_i \sigma^z_i \right)^2 \right\rangle.
\eeq

For the quantum Ising model, we denote by $g=J/\Gamma$ the dimensionless coupling ratio and by $g_c$ its critical value, respectively. This definition of $g$ for the quantum Ising model is chosen in analogy to the $t-V$ model,  such that in both cases the classical Ising model is recovered in the large-$g$ limit. 
For the quantum Ising model, 
the universal properties of the quantum critical point at $g_c$ are described by  the three-dimensional classical Ising model universality class, i.e., the three-dimensional $N=1$ component $\phi^4$-field theory at the Wilson-Fisher fixed point~\cite{Sachdev11}.  As in the case of the $t-V$ model, a line of thermal Ising  transitions terminates at the quantum critical point --- in this case the thermal melting of the ferromagnetic order of the low-$\Gamma$ regime. 
To the best of our knowledge, the details of the thermal phase boundary in the vicinity of the quantum critical point has not been reported previously, and we thus perform for this purpose QMC simulations employing the SSE algorithm of 
Ref.~\onlinecite{Sandvik03}. 
We can profit for this analysis from the feasibility to perform large-system simulations  close to the quantum critical point in order to assess, e.g., the scaling form in Eq.~(\ref{eq:Tcvsg}). We thus begin our analysis in the following section on the quantum Ising model.

\section{Quantum Ising model }~\label{sec:ising}
\begin{figure}[t]
\includegraphics[width=\columnwidth]{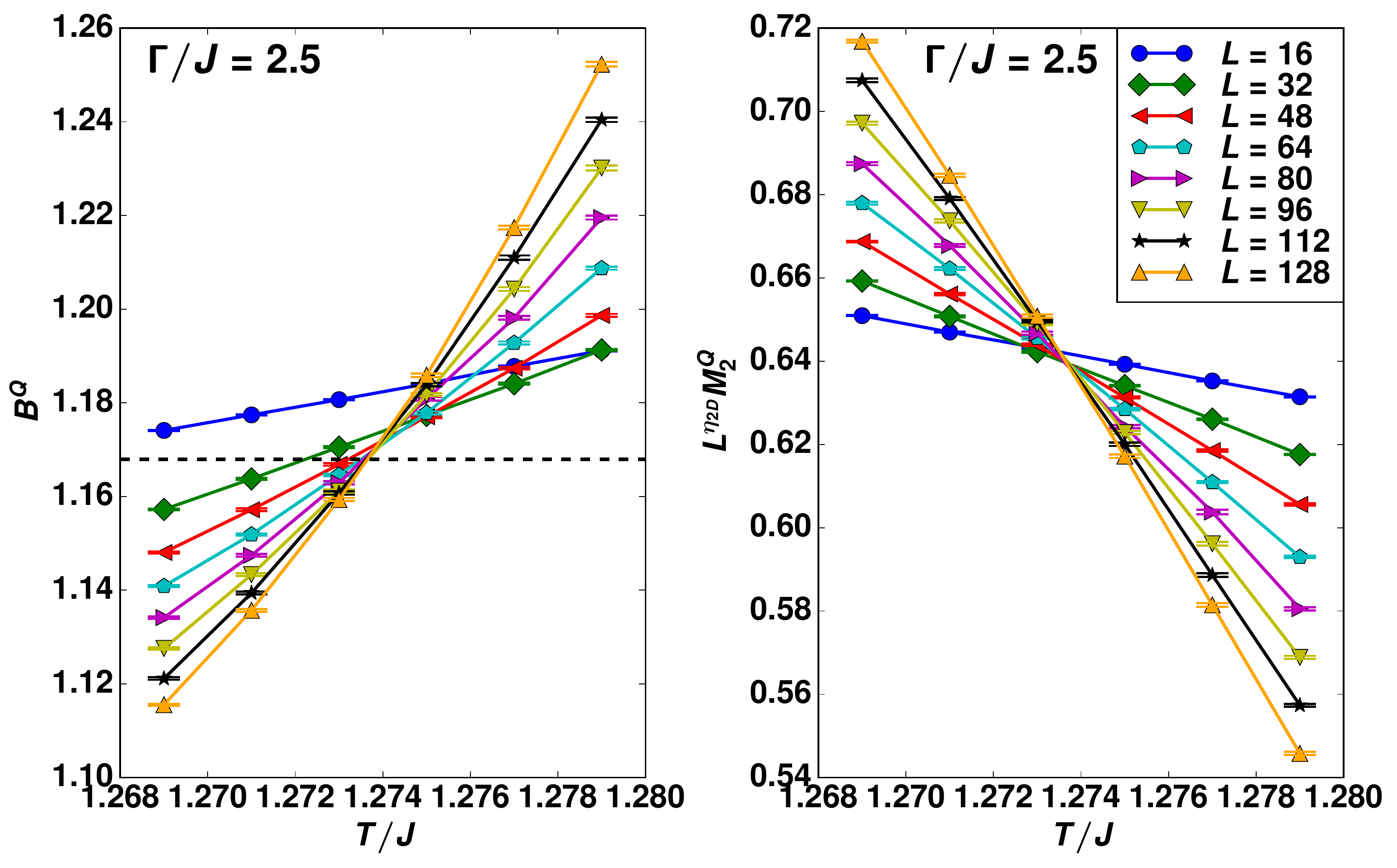}
\caption{(Color online) Finite size data for $B^Q$ (left) and $L^{\eta_{2D}}M_2^Q$ (right) for the quantum Ising model at $\Gamma/J=2.5$ in the critical region of the thermal Ising transition. The dashed line indicates the value of 
the critical Binder ratio for the  classical  Ising model ($\Gamma=0$) on the square lattice.}
\label{fig:BMQraw}
\end{figure}

In order to determine the finite-temperature phase diagram of the quantum Ising model, we performed SSE simulations on systems with periodic boundary conditions of linear size $L$, and $N_s=L^2$ sites, with $L$ up to to 128 at various values of $\Gamma$, focusing on the vicinity of the quantum critical point, as detailed below. At a given fixed value of $\Gamma<\Gamma_c$, we   use standard finite-size scaling analysis to locate the critical temperature. In particular, we can use for this purpose the exactly known critical exponents of the two-dimensional Ising model from the Onsager solution~\cite{Onsager44}: $\nu_{2D}=1$, $\eta_{2D}=1/4$, and $\beta_{2D}=1/8$. Within the critical regime of the 
finite-temperature Ising transition, the order parameter estimator  $M^Q_2$ then follows the leading finite-size scaling form 
\beq
\label{eq:m4scaling}
M_2^Q=L^{-\eta_{2D}}f(t_r L^{1/\nu_{2D}}),
\eeq
in terms of the reduced temperature $t_r=(T-T_c)/T_c$ and the scaling function $f$. 
In addition to $M_2^Q$, one may also consider  the dimensionless Binder ratio~\cite{Binder81}
\beq
\label{eq:bIsing}
B^Q=\frac{M^Q_4}{(M^Q_2)^2},
\eeq
defined in terms of $M^Q_2$ and  the quartic order parameter estimator
\beq
\label{eq:m4Ising}
M^Q_4= \left\langle \left( \frac{1}{N_s} \sum_i \sigma^z_i \right)^4 \right\rangle,
\eeq
with the leading finite-size scaling form
\beq
B^Q=f_B(t_r L^{1/\nu_{2D}}),
\eeq
which is often employed to locate the critical point, in particular if the critical exponents for the phase transition are not known. In fact, the above leading scaling form implies that right at the critical temperature ($t_r=0$), the finite-size data of $B^Q$ for different system sizes $L$  intersect, while for $M^Q_2$ the intersection at $T=T_c$ occurs for the appropriately rescaled data $L^{\eta_{2D}}M_2^Q$.
However, corrections to this leading finite-size scaling  form lead to a systematic drift in the crossing points of the data for successively larger system sizes. This is  seen also in the data in Fig.~\ref{fig:BMQraw}, for the case of $\Gamma/J=2.5$, where we plot 
the finite-size data of both quantities across the thermal critical region for various values of $L$. 
Also indicated in  Fig.~\ref{fig:BMQraw} is the value $1.16793(1)$ of the critical Binder ratio for the classical Ising model ($\Gamma=0$) from  Ref.~\onlinecite{Kamieniarz92}. Closer inspection of the figure shows that the crossing points in the interpolated Binder ratio data for the largest considered system sizes already stabilize at this asymptotic value. 
Our data are thus consistent with the expectation, that $B^Q$  takes
on this thermodynamic limit value along the thermal transition line also for finite values of $\Gamma$. 
%
%
In the vicinity of the quantum critical point, i.e.,  for $\Gamma$ close to $\Gamma_c$, we observe an enhanced impact of scaling corrections, reflecting the fact that the regime of classical scaling surrounding the thermal phase transition line narrows close to the quantum critical point. 
%
\begin{figure}[t]
\includegraphics[width=\columnwidth]{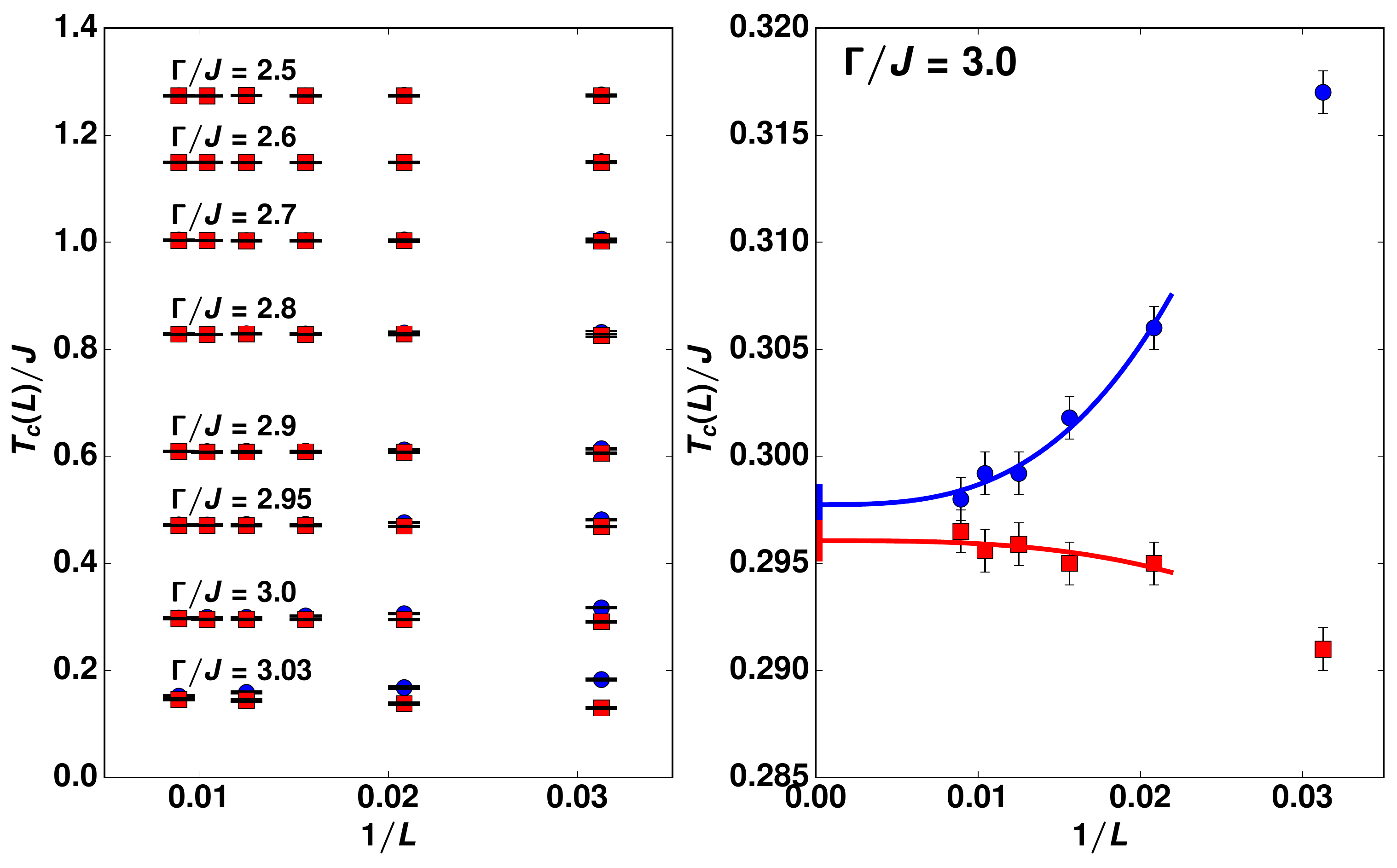}
\caption{(Color online) (Left) Successive crossing points for system sizes $L$ and $L+\Delta L$ in the quantities $B^Q$ (circles) and $M^Q_2$ (squares) as functions of $1/L$ for $\Delta L=16$, as obtained from simulations of the quantum Ising model at different values of $\Gamma$. (Right) Closeup for $\Gamma/J=3.0$. }
\label{fig:BMQcross}
\end{figure}
%
To account for such  scaling corrections we determine, based on the data shown in  Fig.~\ref{fig:BMQraw}, the location of the crossing points  between the data for system sizes $L$ and $L+\Delta L$, and plot the obtained crossing points $T_c(L)$ as functions of $1/L$ in for both quantities in  
Fig.~\ref{fig:BMQcross}, where we employed an offset  $\Delta L=16$.  We find that (i) the Binder ratio $B^Q$ data exhibits larger finite-size drifts in the crossing points than does $L^{\eta_{2D}}M_2^Q$, (ii) the finite-size drifts increase when $\Gamma$ approaches closer to the quantum critical point, and (iii) the crossing points in the Binder ratio $B^Q$ data and for $L^{\eta_{2D}}M_2^Q$ approach the limiting values from opposite sites, thus providing an estimate of  the critical temperature within the temperature window bound by the crossing points for the largest system size. A more  refined estimate of $T_c$ can  be obtained  using the leading finite-size scaling behavior of $T_c(L)$ that describes for fixed $\Delta L$ the convergence of the crossing points towards the critical temperature $T_c$ in the scaling limit~\cite{Qin15},
\begin{equation}
T_c(L)-T_c\propto L^{-1/\nu_{2D}-\omega_{2D}},
\end{equation}
where the dominant irrelevant exponent for the two-dimensional Ising universality class takes on the value $\omega_{2D}=2$ (cf. Ref.~\onlinecite{Pelissetto02}). The right panel illustrates the results from a fit  to these scaling forms at $\Gamma/J=3$, i.e., close to the quantum critical point. The extrapolated values of $T_c$  from both quantities agree within their statistical uncertainty: $T_c/J=0.2977(9)$  is obtained from the Binder ratio $B^Q$ and $T_c/J=0.2960(8)$
from $M^Q_2$.  
We performed the  same analysis for various values of $g$ in order to determine the thermal phase boundary of the ferromagnetic regime. 

\begin{figure}[t]
\includegraphics[width=\columnwidth]{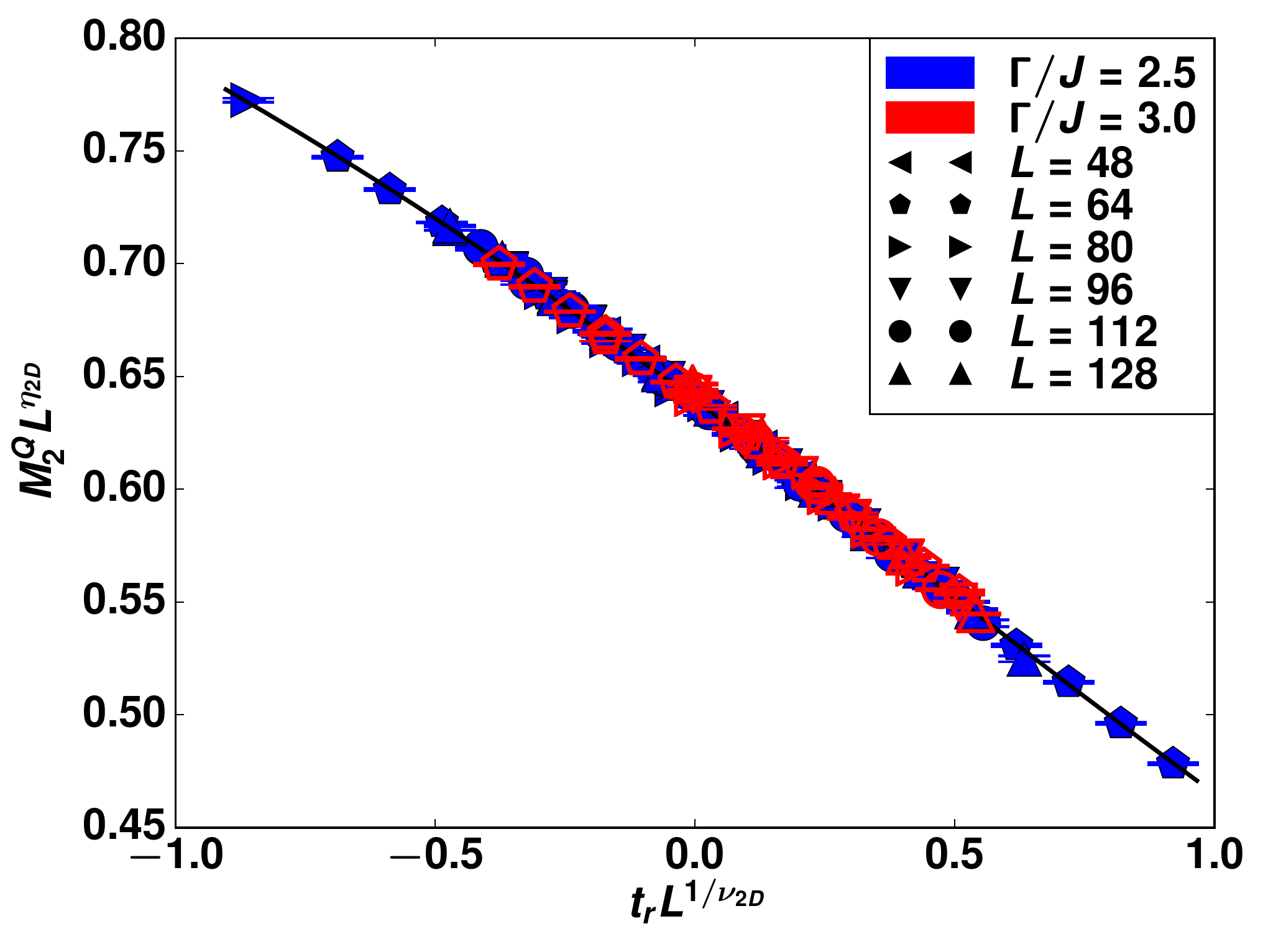}
\caption{(Color online) Data collapse plot of the data for $M^Q_2$ of the quantum Ising model at $\Gamma/J=2.5$ (full symbols). The  line denotes the expanded scaling function. Also included are linearly rescaled data for $M^Q_2$ at $\Gamma/J=3$ (open symbols), as detailed in the text.}
\label{fig:MQcollapse}
\end{figure}

We also compared our results from the crossing point extrapolation to a standard  data collapse analysis, based on the finite-size scaling form of $M^Q_2$ in Eq.~(\ref{eq:m4scaling}). 
This allows us to extract an approximation for $T_c$, since  it enters via the scaling ansatz in terms of the reduced temperature $t_r$.
For this purpose, the scaling function $f$ is expanded up to forth order in its argument and we use the Levenberg-Marquardt scheme to fit the finite-size data to the scaling ansatz. To obtain reliable errors on the fit parameters, we performed a   bootstrap sampling during the fitting procedure. In addition to finite-size restrictions, corrections to scaling arise if the difference to the critical temperature becomes too large. In particular, to remain within the scaling regime, the condition $|t_r|L^{1/\nu}\ll 1$ on the argument of the scaling function should be fulfilled, which also justifies the expansion of the scaling function. Nevertheless, in practice, the scaling ansatz is often found reasonable 
up to $|t_r|L^{1/\nu}\sim \mathcal{O}(1)$. 
From our analysis of the shifting crossing points, cf. Fig.~\ref{fig:BMQcross}, we expect a good data collapse based on 
Eq.~(\ref{eq:m4scaling})
for values of $\Gamma$ sufficiently below the quantum critical point, such that subleading finite-size corrections to the scaling ansatz do not prevail within the range of the available system sizes. As an example, we show the result of such a fitting procedure on the data for $\Gamma/J=2.5$ in Fig.~\ref{fig:MQcollapse}. Even though the data collapse appears   satisfactory, we obtain from the bootstrap analysis a mean value for $\chi^2/\mathrm{d.o.f.}\approx 3.5(5)$, which is slightly larger than what may have been expected and hints at remaining finite-size effects. To account for the corresponding systematic errors in a  more quantitative way, we examined the shift of $T_c$ upon varying the minimum system size used in the fitting procedure, and arrive this way at a final estimate of the critical temperature at $\Gamma/J=2.5$ of $T_c/J=1.27369(5)$, which compares well to the estimates from the extrapolated crossing points, $T_c/J=1.2735(7)$ (based on $B^Q$), and $T_c/J=1.2736(6)$ (based on $M^Q_2$). While the data collapse method can thus in principle provide
rather  accurate estimates of $T_c$, care has to be taken with regards to the subleading finite-size effects that become even more pronounced closer to the quantum critical point. For $\Gamma/J=3$, we obtain an estimate of $T_c/J=2.95(1)$ upon varying the fitted system sizes. Even with enhanced finite-size effects, this estimate is still in reasonable agreement with the crossing point analysis discussed above. 
As anticipated from universality, the 
scaling functions obtained from the data collapses within the critical regime at different values of $\Gamma$ are furthermore equal up to a linear rescaling of their arguments and a global prefactor. This is illustrated in Fig.~\ref{fig:MQcollapse}, where the data for $\Gamma/J=3$ are plotted in the same figures as the data for $\Gamma/J=2.5$, after having performed  such a linear rescaling of the $\Gamma/J=3$ data (i.e., $t_r L^{1/\nu_{2D}} \rightarrow a\times t_r L^{1/\nu_{2D}}$ and $L^{\eta_{2D}}M_2^Q \rightarrow  b\times L^{\eta_{2D}}M_2^Q$, with $a=0.1570$ an $b=2.9985$). 

\begin{figure}[t]
\includegraphics[width=\columnwidth]{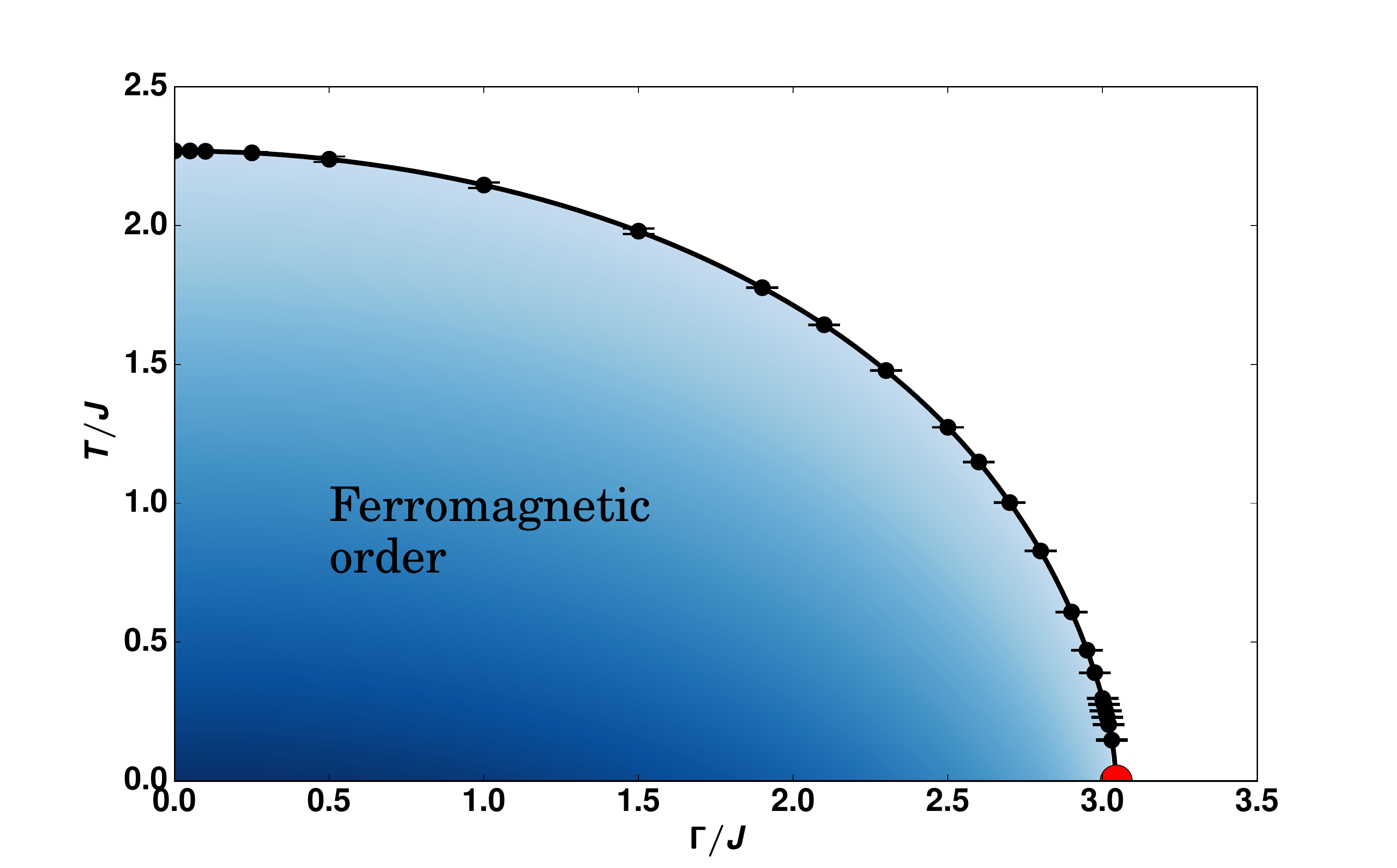}
\caption{(Color online) Thermal phase diagram of the quantum Ising model on the square lattice.}
\label{fig:Qphasediag}
\end{figure}

\begin{figure}[t]
\includegraphics[width=\columnwidth]{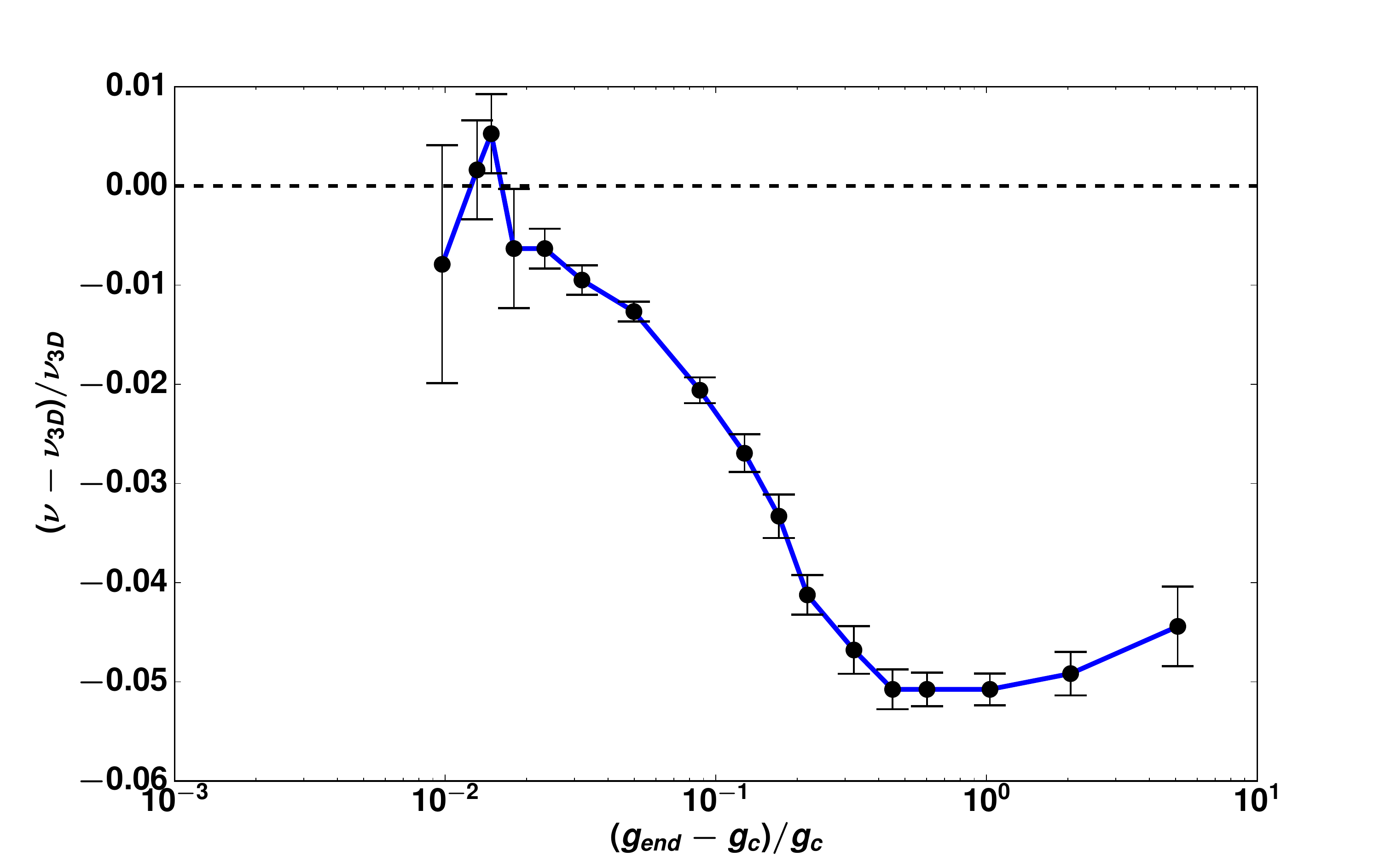}
\caption{(Color online) Deviation of the effective exponent $\nu$ to the value of $\nu_{3D}$ as a function of the relative size of the fit window for the critical line in the quantum Ising model close to the quantum critical point.}
\label{fig:QTcdeviation}
\end{figure}

Based on our combined analysis of crossing points and data collapses, we construct the thermal phase diagram of the quantum Ising model, shown in Fig.~\ref{fig:Qphasediag}. In the limit of vanishing transverse field, $\Gamma\rightarrow 0$, the critical line flattens and $T_c$ tends towards the critical temperature of the classical two-dimensional Ising model~\cite{Onsager44}, $T^{2D}_c/J=2/\ln(1+\sqrt{2})\approx 2.269185$. Close to the quantum critical point, we instead observe an enhanced dependence of $T_c$ on the detuning of $\Gamma$ from the quantum critical value. 
%
\begin{figure}[t]
\includegraphics[width=\columnwidth]{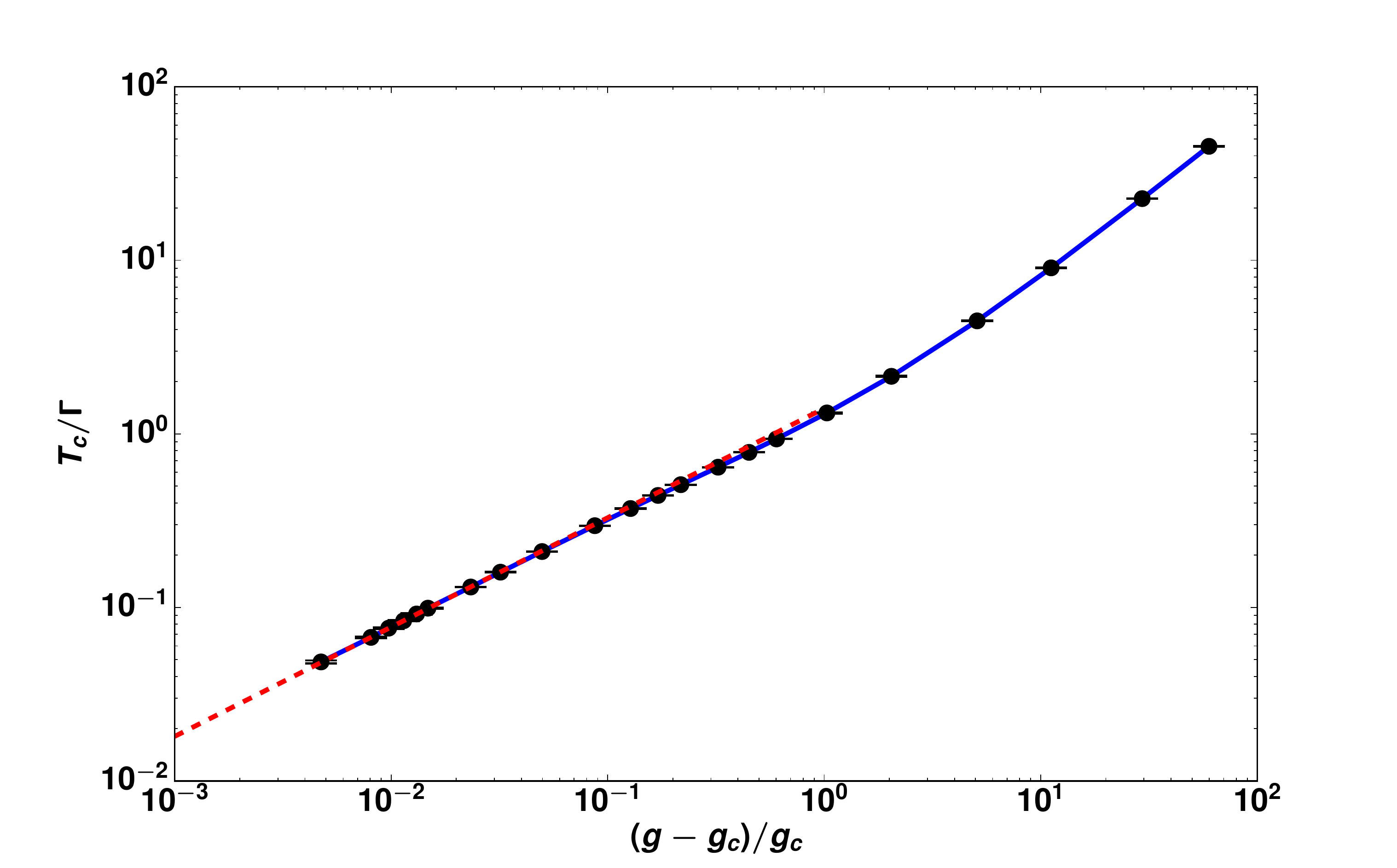}
\caption{(Color online) Scaling of the thermal transition temperature $T_c$ in the vicinity of the quantum critical point.
The dashed line indicates the asymptotic scaling according to Eq.~(\ref{eq:Tcvsg}), with $z=1$ and $\nu=\nu_{3D}$.}
\label{fig:QTcscaling}
\end{figure}
%
To assess, if the asymptotic scaling of $T_c$ in the vicinity of the quantum critical point indeed follows the scaling form of Eq.~(\ref{eq:Tcvsg}), we show in Fig.~\ref{fig:QTcscaling} the dependence of $T_c$ on the relative distance $(g-g_c)/g_c$ to the quantum critical point on a logarithmic scale. We observe a weak, systematic variation in the slope of this curve, with a trend towards an asymptotic scaling in accord with Eq.~(\ref{eq:Tcvsg}) and values of $z=1$ and $\nu=\nu_{3D}=0.62998(3)$, taken from Ref.~\onlinecite{El-Showk14}, as indicated by the dashed line in Fig.~\ref{fig:QTcscaling} (where the quoted recent estimate of $\nu_{3D}$ is based on the conformal bootstrap method~\cite{El-Showk14}).
Furthermore,
the crossover to the limiting large-$g$ behavior $T_c \propto J \propto g$ of the classical Ising model  take place for a relative detuning $(g-g_c)/g_c\approx 1$ of $g$ from the quantum critical coupling strength. 
For a more quantitative assessment of the asymptotic scaling form of $T_c$ near $g_c$, we plot in Fig.~\ref{fig:QTcdeviation} the relative deviation to $\nu_{3D}$ of the value of $\nu$ that results from  
fitting the values of $T_c$ to the scaling from in Eq.~(\ref{eq:Tcvsg}) with $z=1$, depending on the end point 
$g_\text{end}$ of the fit window $[ g_c,g_\text{end}]$. Within the uncertainty set by the fitting procedure, 
we find that (i) the effective exponent $\nu$ indeed approaches the anticipated value  $\nu_{3D}$ upon narrowing the fit window
towards $g_c$,  (ii) for a fit window of relative width $(g_\text{end}-g_c)/g_c\approx 1$, an error of about $5\%$ results in the estimate of $\nu$, and (iii) a relative deviation below  $2\%$ in the estimate for $\nu$ is achieved  for a  relative width of the fit window below about $10\%$.  We can thus extract from the form of the thermal phase transition line in the vicinity of the quantum critical point a reasonable estimate for the critical exponent $\nu$ for this quantum phase transition.

\begin{figure}[t]
\includegraphics[width=\columnwidth]{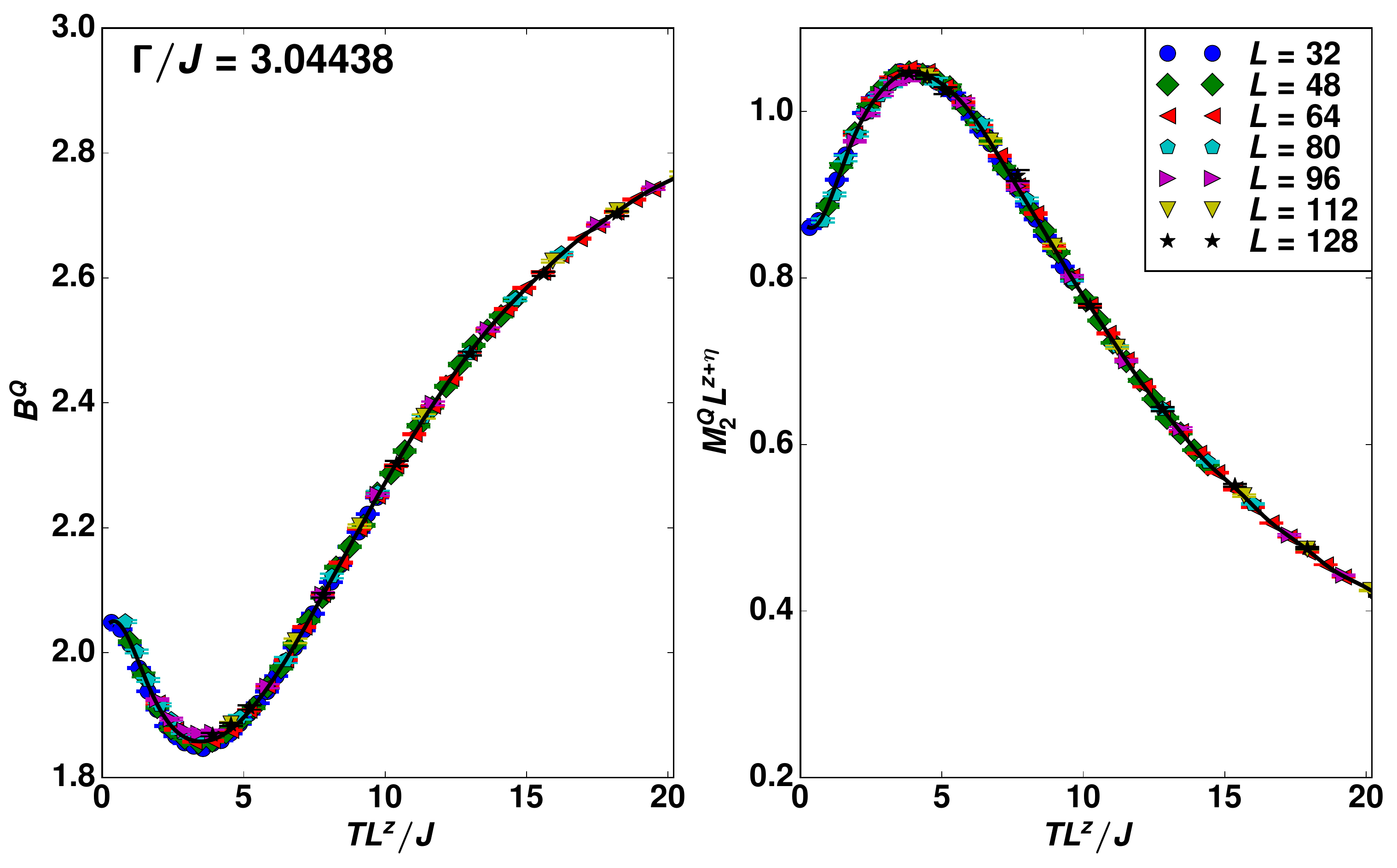}
\caption{(Color online) Data collapse of the finite temperature data for $B^Q$ (left) and  $M^Q_2$ (right) of the quantum Ising model atop the quantum critical point. The lines denote the scaling functions $\tilde{f}_B$ and $\tilde{f}$ respectively,  obtained from a Gaussian process regression.}
\label{fig:Qgcscaling}
\end{figure}

In addition to the shape of the thermal phase transition line, we furthermore examine the finite-temperature scaling within the quantum critical regime. This is accessed most conveniently upon probing the finite-temperature correlations right at $g=g_c$, i.e. atop the quantum critical point. In fact,
the general finite-temperature, two-parameter scaling form for $M^Q_2$ then reduces to~\cite{Melko08} 
\begin{equation}
M^Q_2=L^{-z-\eta}\tilde{f}(T L^z),
\end{equation}
while for the Binder ratio, the scaling form is given as
\begin{equation}
B^Q=\tilde{f}_B(T L^z).
\end{equation}
We can employ the above scaling form of the Binder ratio data to confirm the anticipated value of $z=1$ by a fitting procedure.
However, in contrast to the  data collapse analysis discussed above, the scaling function $\tilde{f}_B$ cannot be represented well by polynomial expansions (cf. Fig.~\ref{fig:Qgcscaling}). We therefore solved the optimization problem for extracting the exponent $z$ through Bayesian inference, which allows us to fit parameters of unknown continuous functions by Gaussian process regression~\cite{Harada11}.  The errors on the fit parameters are again estimated  using bootstrap sampling and varying the initial parameter estimates. To locate the minimum of the log-likelihood function, we use the 
Newton conjugate gradient algorithm. 

The left panel of Fig.~\ref{fig:Qgcscaling} shows the  resulting collapse plot of the Binder ratio data for various system sizes and temperatures, taken at the quantum critical coupling ratio. 
We obtain from this analysis an estimate of $z=1.0027(24)$, well in accord with the  anticipated value of $z=1$.

Fixing thus $z=1$, we next
use the above scaling form for $M^Q_2$ to perform a similar data collapse analysis to extract  the anomalous exponent $\eta$ of the underlying quantum critical point. 
The right panel of 
Fig.~\ref{fig:Qgcscaling} shows the corresponding collapse of the $M^Q_2$ data for various system sizes and temperatures, also taken at the quantum critical coupling ratio. We obtain from this analysis an estimate of $\eta=0.0367(10)$, which agrees well with the expected value of the three-dimensional Ising universality class, $\eta_{3D}=0.03630(2)$, which is again taken from Ref.~\onlinecite{El-Showk14}. The data in Fig.~\ref{fig:Qgcscaling}  clearly demonstrates  the difficulty of approximating the scaling functions $\tilde{f}_B$  and $\tilde{f}$ by low-order polynomial expansions,  as alluded to above. In particular, we find the scaling function $\tilde{f}$ to exhibit a maximum, which implies that for a given finite system, the order parameter estimate shows an initial increase upon increasing the  temperature, starting from the ground state (there is correspondingly a minimum in $\tilde{f}_B$). Note, that this  behavior is however purely a finite-size effect, as the order parameter scales to zero at any finite temperature for $g=g_c$, as well as in the ground state. In fact, the temperature scale set by the maximum in $\tilde{f}$ scales proportional to $L^{-1}\rightarrow 0$ in the thermodynamic limit. The observed maximum  suggests that increasing  the temperature for a fixed system size at $g=g_c$ reduces the effect of the transverse field terms and  thereby favours magnetic ordering, which increases the order parameter estimator $M^Q_2$. Only when the temperature rises beyond the scale set by the maximum in $\tilde{f}$ will the competing thermal fluctuations  eventually reduce the magnetic order. As already mentioned, this effect however disappears completely  in the thermodynamic limit, where critical fluctuations prevail in the ground state. Finally, since for a given finite temperature at $g=g_c$ the systems is not ordered, the values of the Binder ratio $B^Q$ in the thermodynamic limit approaches a value of $3$, and $M_2^Q$ approaches $0$, which also set the limiting values  of the scaling functions $\tilde{f}$ and $\tilde{f}_B$ for large values of their argument, in accord with the behavior seen in  Fig.~\ref{fig:Qgcscaling}.

To summarize this section on the quantum Ising model, we find that for the considered scenario of a conventional Ising quantum critical point, we can accurately determine the finite-temperature thermal transition line and the order parameter fluctuations atop the quantum critical point. The obtained data fits well to  universal scaling forms, allowing us to estimate the values of the critical exponents of the underlying quantum critical point. 
Motivated by these observations, we next turn to the case of the $t-V$ model, and perform a similar analysis based on the recent advances in QMC algorithms to explore the finite-temperature scaling properties near a chiral Ising quantum critical point.

\section{Honeycomb lattice $t-V$ model}~\label{sec:tV}

For the CT-INT simulations of the $t-V$ model on the honeycomb lattice,  a triangular lattice with a two-site unit cell, we employ finite lattices with a rhombus shape.
We considered  lattices of linear size $L$ and $N_s=2L^2$ lattice sites, with $L$ a multiple of $3$ in order to ensure that the Dirac points, which characterize the 
low-energy physics of the half-filled system, are  included in the discretized reciprocal lattice when employing periodic boundary conditions.  
Using the CT-INT algorithm we were able to access values of $L$ up to 21 at temperatures $T$ down to $0.06t$ in the vicinity of the quantum critical point.  

From an analysis of the Binder ratio $B$ and the rescaled quadratic order parameter estimator $L^{\eta_{2D}}M_2$ at various values of $V/t$, we can extract the thermal transition line, similarly to the analysis performed in the previous section. However, in the present case, we are restricted in the range of accessible system sizes, which will be seen to limit our precision as compared to the case of the quantum Ising model.
%
\begin{figure}[t]
\includegraphics[width=\columnwidth]{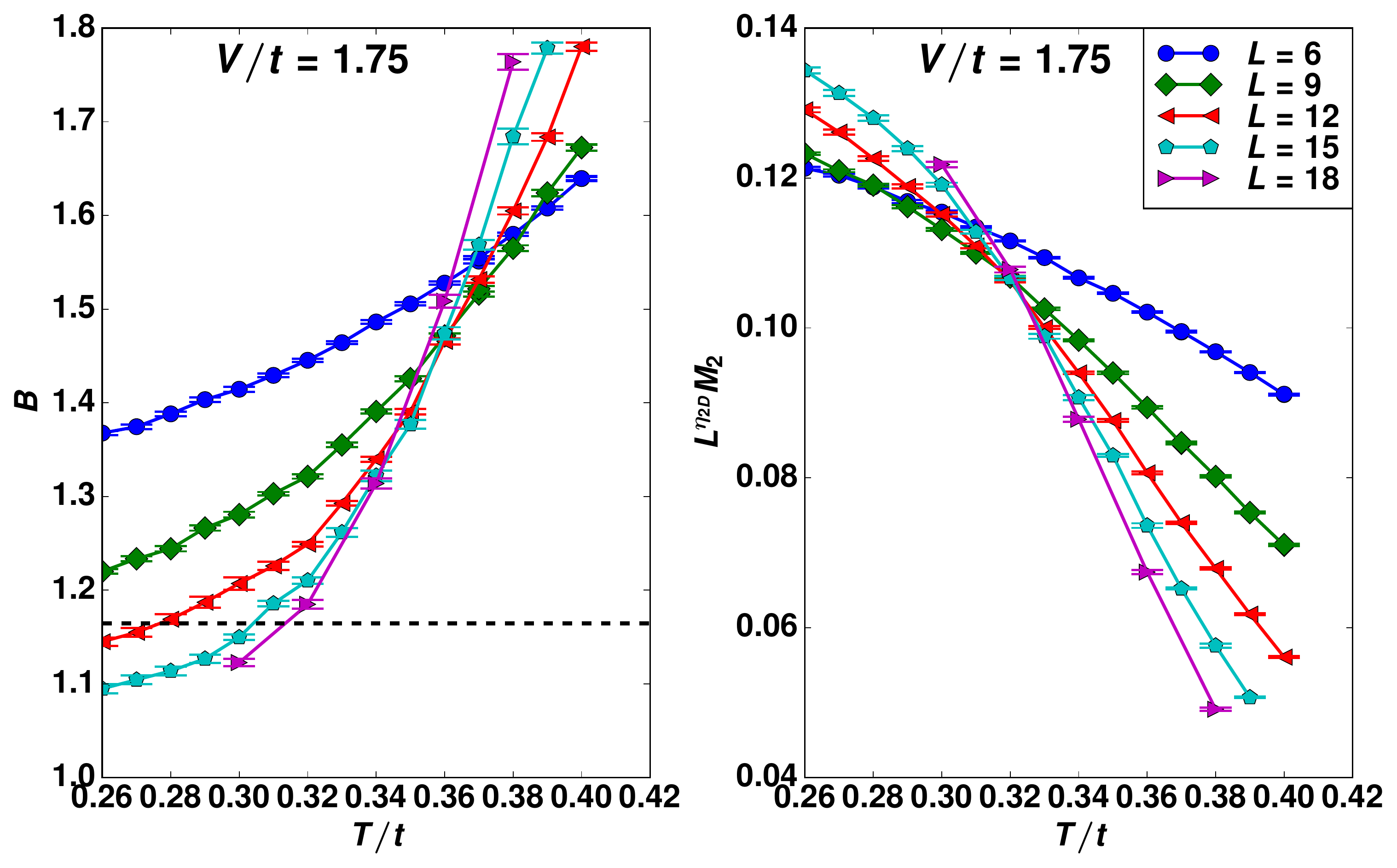}
\caption{(Color online) Finite size data for $B$ (left) and $L^{\eta_{2D}}M_2$ (right) for the spinless fermion $t-V$ model  at $V/t=1.75$ in the critical region of the thermal Ising transition. The dashed line indicates the value of 
the critical Binder ratio for the Ising model on the honeycomb lattice.}
\label{fig:tVBM2}
\end{figure}
%
As an example of our analysis, we show in 
Fig.~\ref{fig:tVBM2} the CT-INT results at a value of $g=V/t=1.75$. As for the case of the quantum Ising model, we observe a systematic drift of the crossing points upon increasing the system size. We again find that both quantities approach convergence from opposite directions, which allows us to bound the transition temperature in the thermodynamic limit. 
 Due to the limited range of accessible system sizes, we are not in a position to employ the asymptotic scaling law for the crossing points, and thus estimate the thermal transition temperatures for the $t-V$ model based on the interval of the bounding crossing points from the largest available system sizes. From the data shown in Fig.~\ref{fig:tVBM2}, we then obtain the estimate $T_c/t=0.325(10)$ for $V/t=1.75$. 
%
\begin{figure}[t]
\includegraphics[width=\columnwidth]{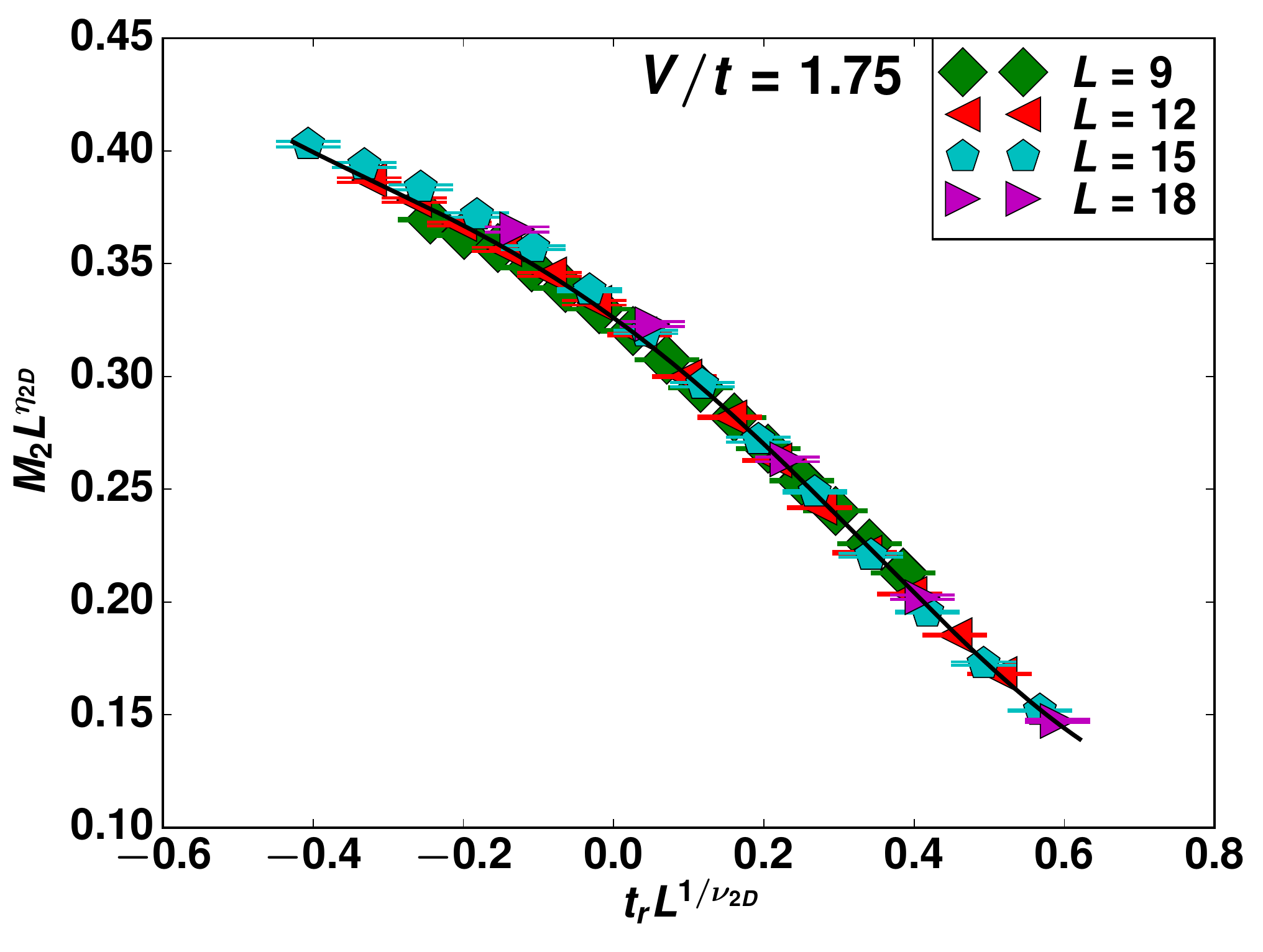}
\caption{(Color online) Data collapse plot of the data for $M$ of the $t-V$ model at  $V/t=1.75$ . The  line denotes the expanded scaling function.}
\label{fig:tVM2dc}
\end{figure}
%
We also employed the data collapse analysis for the $M_2$ data according to 
 \beq
\label{eq:m4scalingtV}
M_2=L^{-\eta_{2D}}f(t_r L^{1/\nu_{2D}}),
\eeq
from which the obtain an estimate of $T_c/t=0.315(20)$ for $V/t=1.75$, with the corresponding data collapse plot shown in Fig.~\ref{fig:tVM2dc}. We also verified, that the scaling function $f$ obtained for the $t-V$ model  can be related by a global rescaling combined with a linear rescaling of the argument to those of the quantum Ising model. 
Figure~\ref{fig:tVBM2}  furthermore includes  the value of the critical Binder ratio for the classical Ising model on the honeycomb lattice, for which we  performed classical Monte Carlo simulations, and which we find to agree with the previous, high precision value $1.1645157(3)$ for the Ising model on the triangular lattice~\cite{Kamieniarz92} (for these simulations, we used the method of Ref.~\onlinecite{Selke07} at the critical temperature 
$T_c/J=1.51865$
of the classical Ising model on the honeycomb lattice~\cite{Fisher67}, keeping the same, rhombus-shaped finite clusters as for the $t-V$ model). 
Our CT-INT data for $B$ in the critical regime are consistent with an approach towards this critical Binder ratio in the thermodynamic limit also for the $t-V$ model in the vicinity of its quantum critical point, even though on the accessible system sizes, the crossing points still reside above this asymptotic value. 
Overall, these observations confirm the anticipated scenario that the thermal transitions out of the CDW phase belong to the universality class of the two-dimensional Ising model for the full range of $V>V_c$.  

\begin{figure}[t]
\includegraphics[width=\columnwidth]{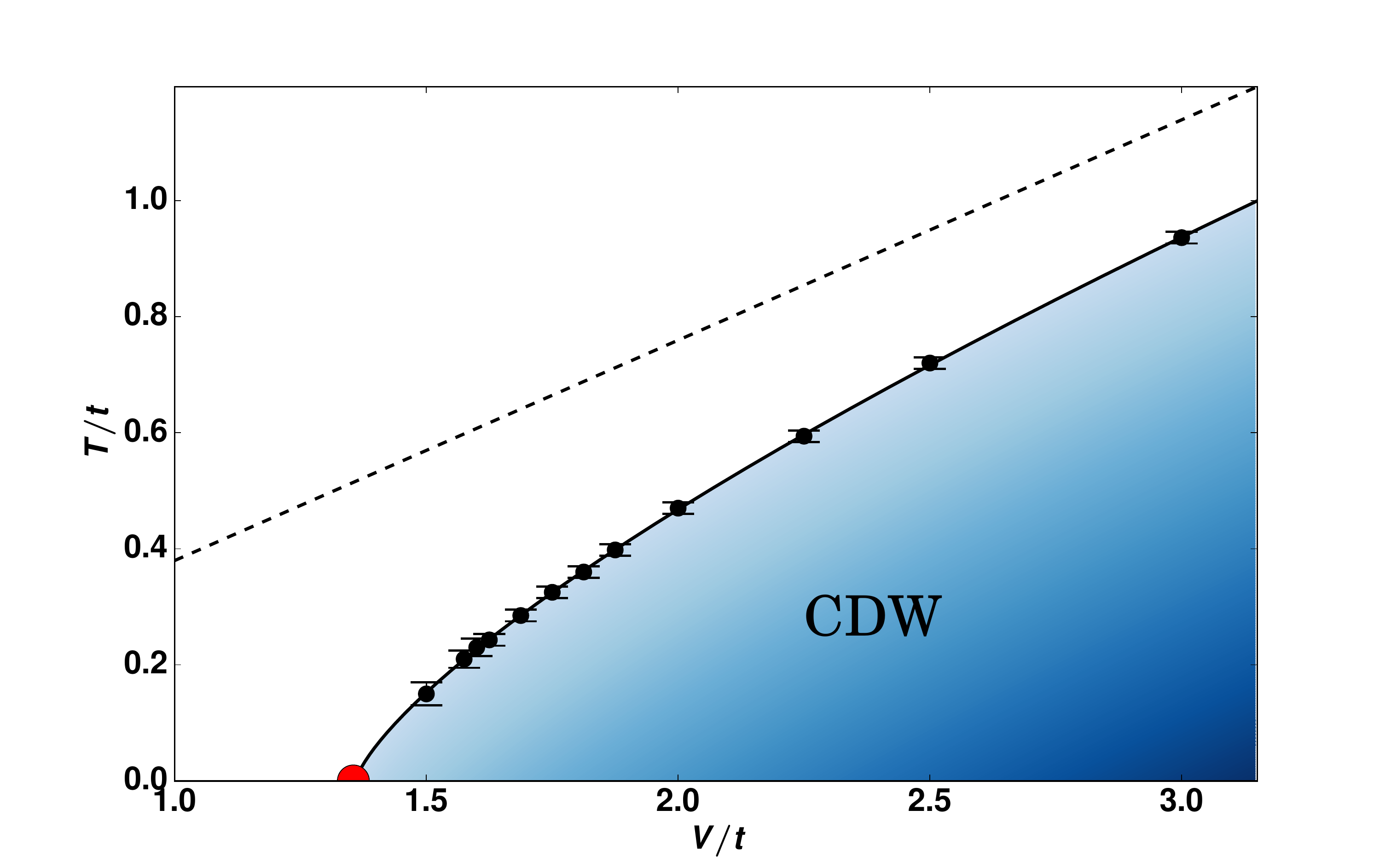}
\caption{(Color online) Thermal phase diagram of the spinless fermion $t-V$  model on the honeycomb lattice. The solid line is a fit of the numerical data to the scaling from Eq.~(\ref{eq:Tcvsg}), and the dashed line indicates the asymptotic scaling of the critical temperature $T_c=0.37966 V$ in the large-$V$  limit. 
}
\label{fig:tVphasediag}
\end{figure}

From an analysis of the CT-INT data for various values of $V/t$, we  eventually obtain the   thermal phase diagram shown in Fig.~\ref{fig:tVphasediag}. In this figure, we also indicated by the dashed line the asymptotic scaling $T_c/V=1.51865/4=0.37966$ that holds in the large-$V$ limit~\cite{Fisher67}. 
%
%
We find that the data within the window $(g-g_c)/g_c<1$ fit well to the scaling form in Eq.~(\ref{eq:Tcvsg}): with $z=1$ from relativistic invariance and leaving $g_c$ and $\nu$ as free parameters, we thereby estimate a value of $g_c=1.359(30)$ for the critical interaction strength and $\nu=0.74(4)$ for the correlation length exponent of the CDW order parameter fluctuations of  the underlying quantum critical point. These values may be compared  to previous estimates based on ground-state QMC calculations:  in Refs.~\onlinecite{Wang14a, Wang15a}, values of $g_c=1.356(1)$ and $\nu=0.80(3)$ were reported (in Ref.~\onlinecite{Wang15a}
the values obtained in Ref.~\onlinecite{Wang14a} were used and checked for consistency, but no independent finite-size analysis was performed), and values of $g_c=1.355(1)$ and $\nu=0.77(3)$  in Ref.~\onlinecite{Li15b}.  
Our results are in overall accord with these previous ground-state estimates, and show
that  the  chiral nature of the underlying quantum critical point affects the scaling of the finite-temperature Ising transition line
in a characteristic way. 

%
\begin{figure}[t]
\includegraphics[width=\columnwidth]{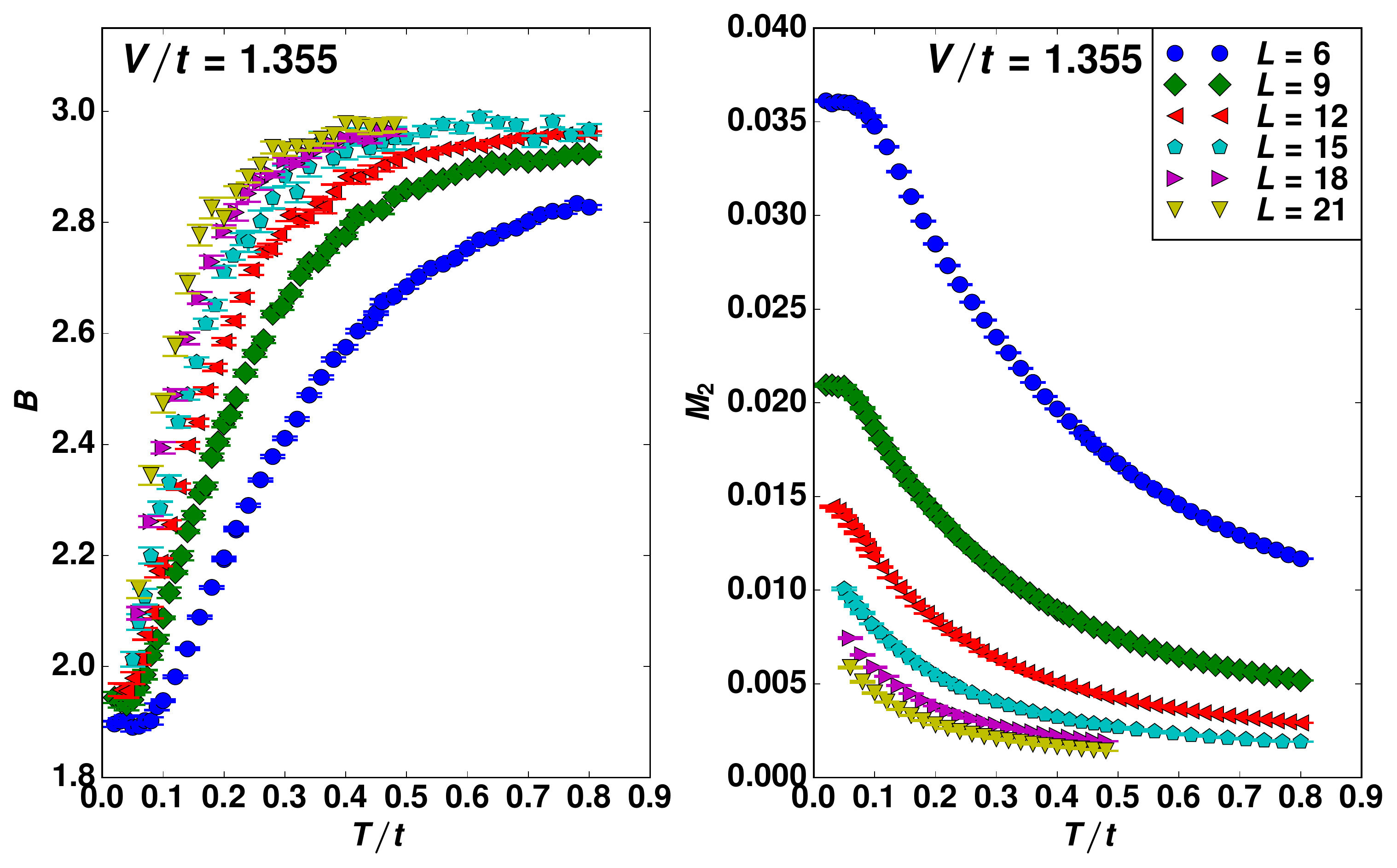}
\caption{(Color online)  Finite-temperature data for $B$ (left) and $M_2$ (right) for the spinless fermion $t-V$ model  at $V/t=1.355$, atop the quantum critical point, for various system sizes.}
\label{fig:tVBM}
\end{figure}
%
To explore further the quantum critical regime, 
we also examined the  scaling of the finite-temperature fluctuations of the CDW order parameter atop the quantum critical point, considering the  value of $g_c=1.355$, taken from Ref.~\onlinecite{Li15b}. The finite-size values for both $B$ and $M_2$ are shown in the left and the right panels of Fig.~\ref{fig:tVBM}, respectively. We consider temperatures up to $T\approx t$, and  significantly larger 
temperatures eventually  extend beyond the universal, quantum critical regime. 
For both $L=6$ and $L=9$, we can clearly identify the low-temperature saturation of both quantities towards their ground state values. For $L=12$, the onset of this saturation can still be seen, while for all larger values of $L$, this saturation apparently happens at a lower temperature scale $\propto 1/L$ than considered here. 
%
\begin{figure}[t]
\includegraphics[width=\columnwidth]{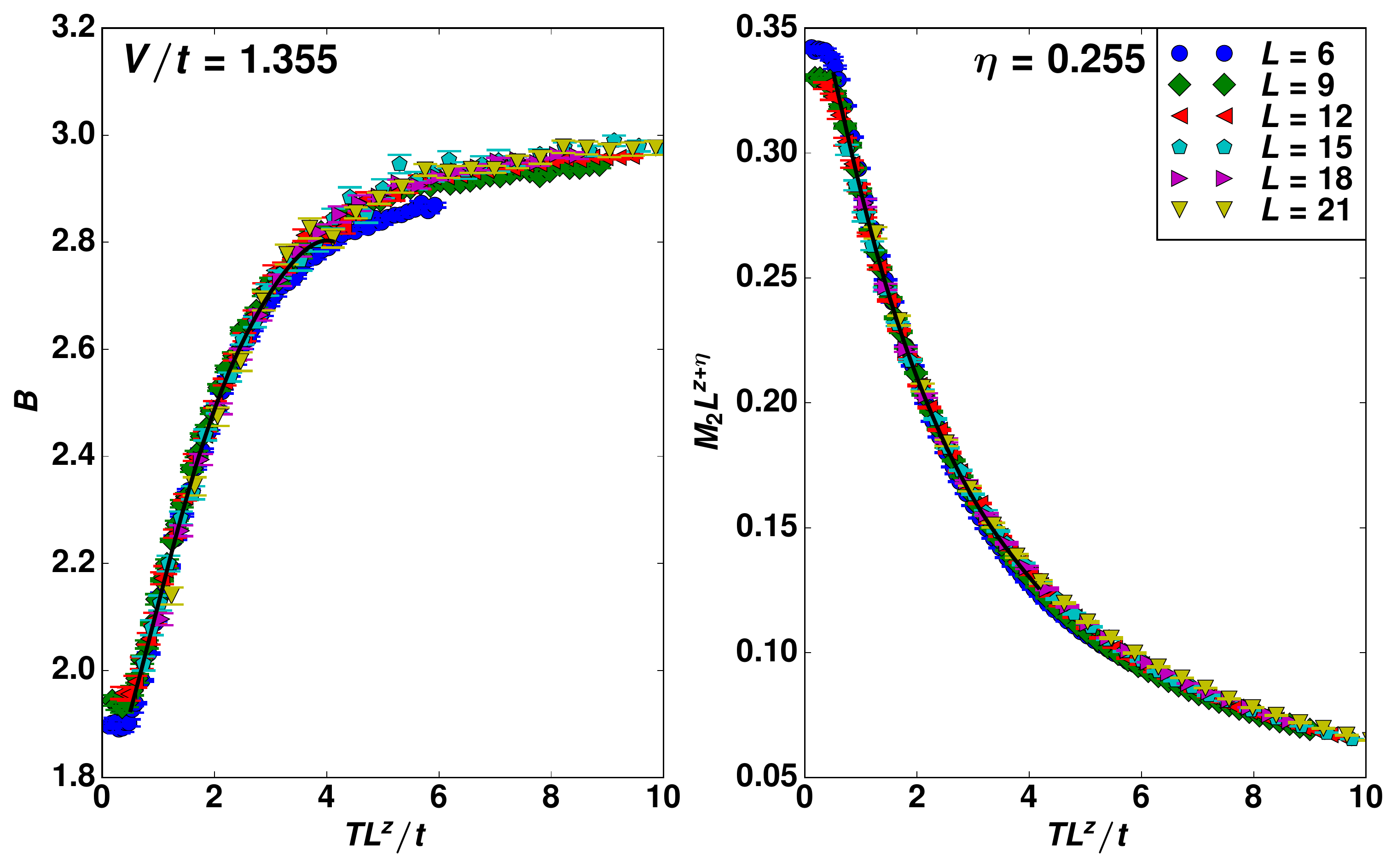}
\caption{(Color online) Data collapse of the finite-temperature data for $B$ (left) and $M_2$ (right) 
 for the spinless fermion $t-V$ model  at $V/t=1.355$, atop the quantum critical point, for various system sizes.
The lines denotes the scaling functions $\tilde{f}_B$ and $\tilde{f}$,
obtained from a Gaussian process regression. 
 }
\label{fig:tVBMsc}
\end{figure}
%
\begin{figure}[t]
\includegraphics[width=\columnwidth]{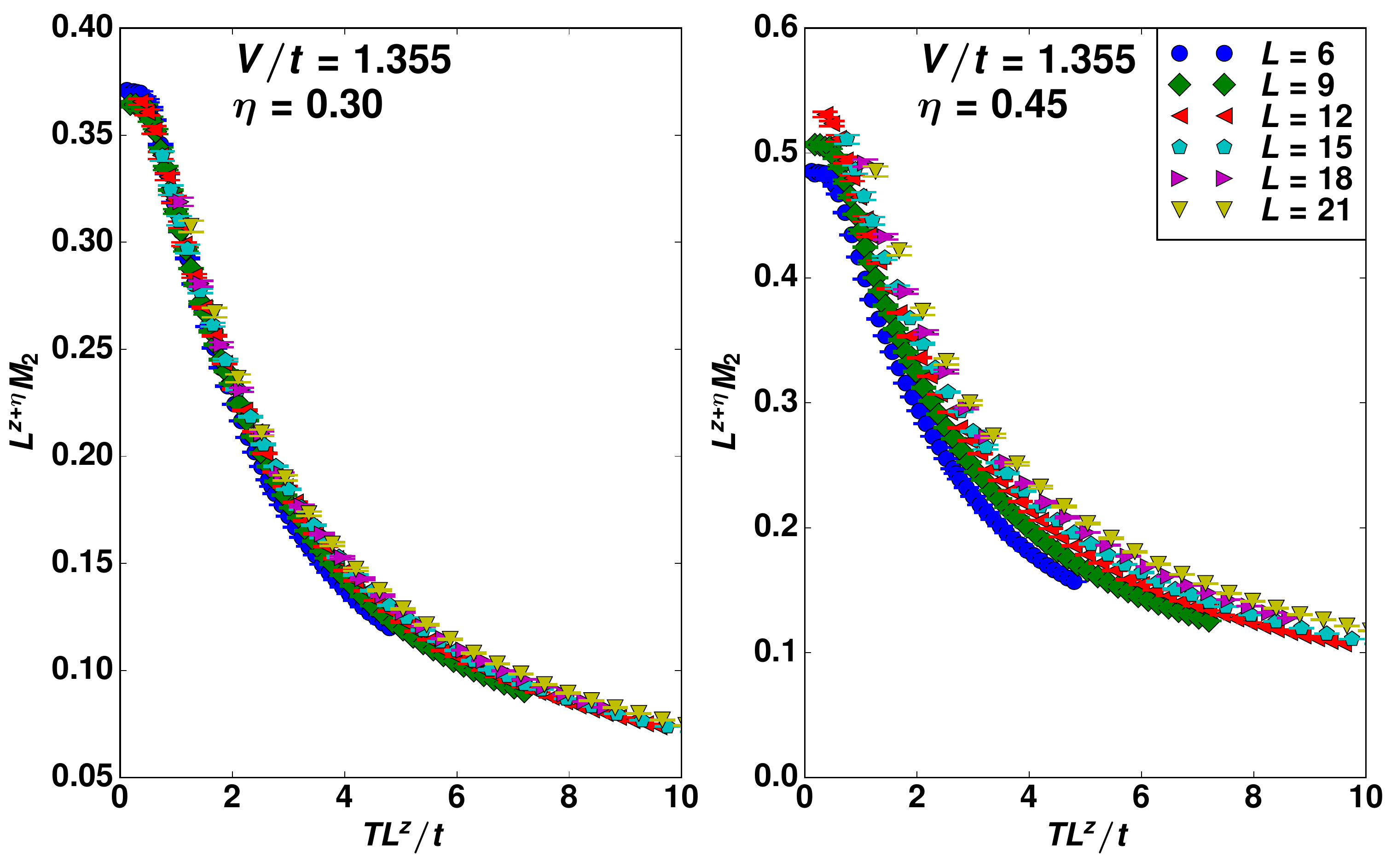}
\caption{(Color online) Attempted data collapse of the finite temperature data  $M_2$ 
 for the spinless fermion $t-V$ model  at $V/t=1.355$, atop the quantum critical point, for various system sizes with 
 $\eta=0.3$~[\onlinecite{Wang14a,Wang15a}] (left) and 
 $\eta=0.45$~[\onlinecite{Li15b}] (right), and $z=1$. 
 }
\label{fig:tVMsc}
\end{figure}
%

For the dimensionless Binder ratio, we expect  a data collapse upon plotting the values of $B$ as a function of $TL^z$, which derives from the scaling form
\begin{equation}
B=\tilde{f}_B(T L^z)
\end{equation}
in the quantum critical regime atop the quantum critical point, i.e., for $g=g_c$, as considered here. 
As shown in the left panel of Fig.~\ref{fig:tVBMsc},
such a data collapse is indeed feasible over the full range of the  finite-temperature data, only the $L=6$ data exhibit  systematic deviations, in particular at low temperatures, indicative of the finite-size saturation effects already mentioned above. Again, the obtained quantum critical Binder ratio data are in accord with an asymptotic approach towards a value of $3$ for large values of $TL^z$. However, the statistical noise in the data for the larger system sizes limits the stability of data fitting procedures in this regime. 
 An unbiased fit of the Binder ratio data, using the Bayesian inference method~\cite{Harada11} applied to the scaling function $\tilde{f}_B$ over the fit range for $TL^z$ from 0.5 to 4 results in a value of $z=0.99(1)$, also indicated 
in Fig.~\ref{fig:tVBMsc}. This result is in accord with the anticipated value of $z=1$.
The good overall collapse of the Binder ratio data indicates that within the accessible range of system sizes and temperatures, we can indeed probe the scaling regime atop the quantum critical point. 

We thus also performed a data collapse analysis of the finite-temperature data for $M_2$, based on its scaling form\begin{equation}
M_2=L^{-z-\eta}\tilde{f}(T L^z),
\end{equation}
at $g=g_c$, 
in terms of the critical exponents $z=1$ and $\eta$. Due to the explicit $\eta$ dependence, we can estimate  the value of this critical exponent upon monitoring the quality of the data collapse for varying values of $\eta$. A good overall data collapse results for $\eta \approx 0.25$, as shown in the right panel of Fig.~\ref{fig:tVBMsc}, with again  the $L=6$ data exhibiting the most pronounced deviations at low temperatures. A fit based on the Bayesian inference method~\cite{Harada11} applied to the scaling function $\tilde{f}$ over the fit range for $TL^z$ from 0.5 to 4 results in a value of $\eta=0.255(10)$, indicated 
in Fig.~\ref{fig:tVBMsc}.

\begin{table}
\begin{ruledtabular}
\begin{tabular}{ | c | c | c | }
 \hline
 Method & $\nu$ & $\eta$ \\ \hline \hline
 $4-\epsilon$, 1st order [\onlinecite{Rosenstein93}] & $0.709$ & $0.577$ \\ \hline 
 $4-\epsilon$, 2st order [\onlinecite{Rosenstein93}] & $0.797$ & $0.531$ \\ \hline 
 FRG (linear cutoff) [\onlinecite{Hoefling02, Rosa01}] & $0.927$ & $0.525$ \\ \hline
 FRG (exp. cutoff) [\onlinecite{Rosa01}] & $0.962$ & $0.554$ \\ \hline
 FRG [\onlinecite{Vacca15}] & $0.929$ & $0.602$ \\ \hline
 $1/N$ expansion [\onlinecite{Hoefling02}] & $0.738$ & $0.635$ \\ \hline
 CT-INT (GS) [\onlinecite{Wang14a}]& $0.80(3)$ & $0.302(7)$  \\ \hline 
 MQMC (GS) [\onlinecite{Li15b}] & $0.77(3)$ & $0.45(2)$ \\ \hline 
 LCT-INT (GS) [\onlinecite{Wang15a}] & $0.80(3)$ & $0.302(7)$  \\ \hline
 CT-INT (finite $T$), here & $0.74(4)$ & 0.275(25) \\ \hline 
\end{tabular} 
\caption{Overview of reported estimates for the critical exponents $\nu$ and $\eta$ for the $N_f=1$, $Z_2$-Gross-Neveu theory in 2+1 dimensions (analytic approaches) as well as the spinless fermion $t-V$ model on the honeycomb lattice (QMC approaches).
GS indicates that  ground state correlations were targeted, while the results reported here (in the last row) were obtained from the  finite-temperature scaling.
Note, that for the LCT-INT  results in Ref.~\onlinecite{Wang15a}, with system sizes up to $L=18$, the values of Ref.~\onlinecite{Wang14a} (with $L$ up to $15$) were used and checked for consistency, but no independent finite-size analysis was performed. In the MQMC  approach of Ref.~\onlinecite{Li15b}, system sizes up to $L=24$ were employed. 
}
\label{table1}
\end{ruledtabular}
\end{table}

We next compare our estimate for $\eta$ to the values obtained from previous QMC simulations, which targeted   ground state correlations. In Refs.~\onlinecite{Wang14a} and \onlinecite{Wang15a}, a value of $\eta=0.307(2)$ was obtained from simulations with  $L$ up to 18. In Ref.~\onlinecite{Li15b}, employing simulations up to $L=24$, a value of $\eta=0.45(2)$ was reported. 
Figure~\ref{fig:tVMsc} shows attempted collapse plots of our finite-temperature data, using  values of $\eta=0.3$ (left) and 
$\eta=0.45$ (right). While for $\eta=0.3$, an acceptable collapse of the data is observed at least for $TL^z<2t$, a value of $\eta=0.45$ does not lead to a good data collapse. While corrections to scaling may still explain the systematic deviations from a perfect collapse seen in the left panel of Fig.~\ref{fig:tVMsc} for $\eta=0.3$, a value of $\eta=0.45$  appears too large:
the fact that the Binder ratio data exhibits a good overall collapse on the accessible system sizes and temperatures (independently of the value of $\eta$) suggests that one can also collapse the available data for $M_2$, given an appropriate value of $\eta$. 
From this analysis, we thus estimate a value of $\eta \approx 0.25-0.3=0.275(25) $ to be in accord with our finite-temperature data inside the quantum critical regime.  

Finally, and on a  more qualitative account, we  note   that for the $t-V$ model the scaling function $\tilde{f}$ does not exhibit a local maximum (cf. the right panel of Fig.~\ref{fig:tVBMsc}), in contrast to the corresponding  scaling function for the quantum Ising model (cf. Fig.~\ref{fig:Qgcscaling}). This observation provides a direct qualitative distinction of the two different universality classes at the corresponding quantum critical points.

\section{Discussion}~\label{sec:discussion}

We employed quantum Monte Carlo simulations to analyzse  thermal Ising transition lines in the vicinity of
quantum phase transitions of two-dimensional quantum lattice models  and related the thermal scaling properties
to the critical properties of the quantum critical point. For the quantum Ising model on the square lattice, our numerical
results for the critical exponents $\nu$ and $\eta$, which we obtained from the scaling of the Ising transition line and
the order parameter correlations in the quantum critical regime, compare well to previous estimates of these critical 
exponents for the three-dimensional Ising universality class. In the case of the spinless-fermion $t-V$ model on the
honeycomb lattice, 
our simulations have shown that it is  feasible to estimate the quantum critical scaling exponents from the finite-temperature scaling and the Ising transition-line, allowing us, e.g., to discern the conventional Ising quantum phase transition in  the quantum Ising model from the chiral Ising transition in the fermionic system.
We obtained values of  $\nu$ and $\eta$, which compare well to previous results based
on ground-state calculations. 
The critical exponents of the associated 
$N_f=1$, $Z_2$-Gross-Neveu theory in 2+1 dimensions
have also  been estimated within  various analytic approaches,  based on field theoretic and resummation techniques. 
Table~\ref{table1} summarizes these various results for the critical exponents (cf. Ref.~\onlinecite{Otsuka15} for a similar compilation of estimates for the critical exponents of various chiral universality classes). For the $\epsilon$-expansion and the $1/N$-expansion, the critical exponent $\nu$ is in reasonable agreement with the quantum Monte Carlo values, while the reported values from the functional renormalization group (FRG) are systematically slightly larger. The anomalous exponent $\eta$ is obtained larger  than the quantum Monte Carlo estimates by all the analytical approaches, with a factor of about two difference to the results of Refs.~\onlinecite{Wang14a,Wang15a} and our finite-temperature findings;  the difference to the value reported in Ref.~\onlinecite{Li15b} is less prominent. In general, one may consider 
an insufficient
approach to the scaling region 
 to plague quantum Monte Carlo estimates of critical exponents. However, the Binder ratio data in our simulations exhibit a good overall relativistic scaling (and independently of the values of $\nu$ and $\eta$), suggesting that 
the critical exponents have also been accessed within the scaling regime.
For the future, it would be desirable to eventually overcome the reported deviations between the various estimates and to eventually reconcile analytic and numeric approaches for the chiral Ising universality class to a similar precision as has been achieved  for the Wilson-Fisher fixed-points.

{\it Note} -- An independent finite-temperature study of the $t-V$ model, based on a hybrid SSE CT-QMC algorithm, has been performed by Wang, Liu, and Troyer~\cite{Wang16}. Whenever there is an overlap, their results are in full agreement to 
those reported here. 

\section*{Acknowledgments}
We thank S. Chandrasekharan, Z. Y. Meng, D. Mesterh\'{a}zy, M. Scherer, M. Troyer, and L. Wang for discussions,
and acknowledge support by the Deutsche Forschungsgemeinschaft (DFG) under grant FOR 1807 and RTG 1995. Furthermore, we thank the IT Center at RWTH Aachen University and the JSC J\"ulich for access to computing time through JARA-HPC.
SW thanks the KITP Santa Barbara for hospitality during the program ``Entanglement in Strongly-Correlated Quantum Matter''. This research was supported in part by the National Science Foundation under Grant No. NSF PHY11-25915.


%
%

\begin{references}


\bibitem{CastroNeto09}
A. H. Castro Neto, N. M. R. Peres, K. S. Novoselov and A. K. Geim, Rev. Mod. Phys. {\bf{81}}, 109 (2009).


\bibitem{Uehlinger13}
T. Uehlinger, G. Jotzu, M. Messer, D. Greif, W. Hofstetter, U. Bissbort, and T. Esslinger, Phys. Rev. Lett. {\bf 111}, 185307 (2013).


\bibitem{Vojta00}
M. Vojta, Y. Zhang and S. Sachdev, Phys. Rev. Lett. {\bf{85}}, 4940 (2000).


\bibitem{Khveshchenko01}
D. V. Khveshchenko and J. Paaske, Phys. Rev. Lett, {\bf{86}}, 4672 (2001).


\bibitem{Hasan10}
M. Z. Hasan and C. L. Kane, Rev. Mod. Phys. {\bf{82}}, 3045 (2010).


\bibitem{Gross74}
D. J. Gross and A. Neveu, Phys. Rev. D {\bf{10}}, 3235 (1974).


\bibitem{Herbut06}
I. F. Herbut, Phys. Rev. Lett. {\bf{97}}, 146401 (2006).


\bibitem{Herbut09a}
I. F. Herbut, V. Juricic and O. Vafek, Phys. Rev. B {\bf{80}}, 075432 (2009).


\bibitem{Rosenstein93}
B. Rosenstein, H.-L. Yu and A. Kovner, Phys. Lett. B {\bf{314}}, 3816 (1993).


\bibitem{Rosa01}
L. Rosa, P. Vitale and C. Wetterich, Phys. Rev. Lett. {\bf{86}}, 958 (2001).


\bibitem{Hoefling02}
F. H\"ofling, C. Nowak and C. Wetterich, Phys. Rev. B {\bf{66}}, 205111 (2002).


\bibitem{Mesterhazy12}
D. Mesterh\'{a}zy, J. Berges, L. von Smekal, Phys. Rev. B {\bf 86}, 245431 (2012).


\bibitem{Vacca15}
G. P. Vacca and L. Zambelli, Phys. Rev. D 91, 125003 (2015).


\bibitem{Stephanov95}
M. A. Stephanov, Phys. Rev. D {\bf{52}}, 3746 (1995).


\bibitem{Sachdev11}
S. Sachdev, Quantum Phase Transition, Cambridge Univ. Press. (2011).


\bibitem{footnote1}
At the upper critical dimension of the $\phi^4$-theory, additional logarithmic corrections to mean-field behavior arise, and have recently been identified in the finite-temperature transition line near 
a corresponding quantum phase transition in a three-dimensional coupled dimer system in Ref.~\onlinecite{Qin15}. 


\bibitem{Janssen14}
L. Janssen and I. F. Herbut, Phys. Rev. B {\bf{89}}, 205403 (2014).


\bibitem{Wang14a}
L. Wang, P. Corboz and M. Troyer, New J. Phys. {\bf{16}}, 103008 (2014).


\bibitem{Li15b}
Zi-Xiang Li, Yi-Fan Jiang and Hong Yao, New J. Phys. {\bf{17}}, 085003 (2015).


\bibitem{Wang15a}
L. Wang, M. Iazzi, P. Corboz and M. Troyer, Phys. Rev. B {\bf{91}}, 235151 (2015).


\bibitem{Huffman14}
E. F. Huffman and S. Chandrasekharan, Phys. Rev. B {\bf{89}}, 111101 (2014).


\bibitem{Li15a}
Zi-Xiang Li, Yi-Fan Jiang and Hong Yao, Phys. Rev. B {\bf{91}}, 241117(R), (2015).


\bibitem{Iazzi15}
M. Iazzi and M. Troyer, Phys. Rev. B {\bf{91}}, 241118(R), (2015).

\bibitem{Wang15b}
L. Wang, Ye-Hua Liu, M. Iazzi, M. Troyer and G. Harcos, arXiv:1506.05349.

\bibitem{Wei16}
Z. C. Wei, C. Wu, Y. Li, S. Zhang, and T. Xiang, arXiv:1601.01994. 

\bibitem{Li16}
Z.X. Li, Y. F. Jiang, and H. Yao, arXiv:1601.05780.


\bibitem{Broecker15}
P. Broecker and S. Trebst, arXiv:1511.02878. 

\bibitem{Sandvik03}
A. W. Sandvik, Phys. Rev. E {\bf{68}}, 056701 (2003).


\bibitem{Chandrasekharan10}
S. Chandrasekharan, Phys. Rev. D {\bf{82}}, 025007 (2010).


\bibitem{Blote02}
H. W. J. Blote and Y. Deng, Phys. Rev. E {\bf{66}}, 066110 (2002).


\bibitem{Onsager44}
L. Onsager, Phys. Rev. {\bf{65}}, 117 (1944).


\bibitem{Binder81}
K. Binder, Z. Phys. B {\bf{43}}, 11940 (1981).


\bibitem{Kamieniarz92} 
G. Kamieniarz and H. W. J. Bl\"ote, J. Phys. A: Math. Gen. {\bf 26}, 201 (1993).


\bibitem{Qin15}
Y. Q. Qin, B. Normand, A. W. Sandvik, and Z. Y. Meng, arXiv:1506.06073 (2015). 


\bibitem{Pelissetto02}
A. Pelissetto and E. Vicari, Phys. Rept. {\bf{368}}, 549 (2002).


\bibitem{El-Showk14}
S. El-Showk, M. F. Paulos, D. Poland, S. Rychkov, D. Simmons-Duffin and A. Vichi, J. Stat. Phys. {\bf{157}}, 869 (2014).


\bibitem{Melko08}
R. G. Melko and R. K. Kaul, Phys. Rev. Lett. {\bf{100}}, 017203 (2008).


\bibitem{Harada11}
K. Harada, Phys. Rev. E {\bf{84}}, 056704 (2011).


\bibitem{Selke07} 
W. Selke,  J. Stat. Mech. P04008 (2007).


\bibitem{Fisher67} 
M. E. Fisher, Rep. Prog. Phys. 30, 615 (1967). 

\bibitem{Otsuka15}
Y. Otsuka, S. Yunoki, and  S.Sorella, arXiv:1510.08593.

\bibitem{Wang16}
L. Wang, Y.-H. Liu, and M.Troyer, Phys. Rev. B {\bf{93}}, 155117 (2016).

\end{references}
\end{document}